\documentclass[a4paper,11pt]{article}
\usepackage[utf8]{inputenc}
\usepackage[normalem]{ulem}
\usepackage[left=2.5cm,right=2.5cm,top=2cm,bottom=2.5cm]{geometry}
\usepackage{epsfig,amsfonts,amssymb}
\usepackage{xcolor}
\usepackage{amsmath,bm}
\usepackage{framed}
\usepackage{varioref}
\colorlet{shadecolor}{lightgray}
\usepackage{hyperref}
\usepackage{mathrsfs}
\usepackage{parskip}
\usepackage{cite}

\def \be {\begin{equation}}
\def \ee {\end{equation}}
\def \bea {\begin{eqnarray}}
\def \eea {\end{eqnarray}}
\def \nn {\nonumber}

\labelformat{equation}{Eq.~(#1)} 

\begin{document}

\baselineskip 24pt

\begin{center}

{\Large \bf Celestial Eikonal Amplitudes in the Near-Horizon Region}

\end{center}

\vskip .6cm
\medskip

\vspace*{4.0ex}

\baselineskip=18pt

\centerline{\large \rm Karan Fernandes$^{a,b}$, Feng-Li Lin$^{a,b}$ and Arpita Mitra$^{c}$}

\vspace*{4.0ex}

\centerline{\large \it ~$^a$Department of Physics, \\
National Taiwan Normal University, Taipei, 11677, Taiwan}

\centerline{\large \it ~$^b$Center of Astronomy and Gravitation, National Taiwan Normal University, Taipei 11677, Taiwan}

\centerline{\large \it ~$^c$ Department  of  Physics,  Pohang  University  of  Science  and  Technology,  Pohang  37673,  Korea}

\vspace*{1.0ex}
\centerline{\small E-mail: karanfernandes86@gmail.com, fengli.lin@gmail.com, arpitamitra89@gmail.com}

\vspace*{5.0ex}

\centerline{\bf Abstract} \bigskip

We investigate the celestial description of an eikonal amplitude for the scattering of massless scalars mediated by soft gravitons in the near-horizon region of a large eternal Schwarzschild black hole. Our construction thus provides a celestial conformal field theory on the horizon corresponding to a non-perturbative scattering process that accounts for event horizons on asymptotically flat spacetimes. From the \emph{known} two-dimensional near-horizon scattering amplitude computed within the effective field theory framework, we first construct a four-dimensional amplitude in a flat spacetime frame around the bifurcation sphere strictly in a small angle approximation limit. While the kinematics of external particles in this frame at leading order are analogous to a Minkowski spacetime, the eikonal amplitude differs from those about flat spacetime due to the near-horizon scattering potential. We construct a celestial correlator following a Mellin transform that provides an all loop order result, with a universal leading ultraviolet (UV) soft scaling behavior of the conformally invariant cross-ratio, and an infrared (IR) pole for the scaling dimension at each loop order. We argue these properties manifest soft graviton exchanges in the near-horizon region and, consequently, the soft UV behavior of the amplitude.

\vfill \eject

\baselineskip 18pt

\tableofcontents

\section{Introduction}

Aspects of a holographic correspondence relating scattering amplitudes with correlation functions of a dual conformal field theory have been recently realized on the celestial sphere at null infinity of asymptotically flat spacetimes~\cite{Cheung:2016iub, Pasterski:2016qvg, Pasterski:2017kqt, Pasterski:2017ylz, Cardona:2017keg,Donnay:2018neh, Stieberger:2018edy,Stieberger:2018onx,Pate:2019mfs,Puhm:2019zbl,Pate:2019lpp, Nandan:2019jas,Fotopoulos:2019vac,Banerjee:2020kaa,Fotopoulos:2020bqj,Gonzalez:2020tpi,Donnay:2020guq,Pasterski:2020pdk, Gonzo:2022tjm,Arkani-Hamed:2020gyp,Atanasov:2021cje,Guevara:2021abz,Atanasov:2021oyu,Strominger:2021lvk,Strominger:2021mtt,Pasterski:2021raf,Pasterski:2021rjz,Raclariu:2021zjz,Crawley:2021ivb,Donnay:2022aba,PipolodeGioia:2022exe,Pasterski:2022lsl,McLoughlin:2022ljp,Casali:2022fro,Mizera:2022sln,Cotler:2023qwh,Donnay:2023kvm}. This correspondence follows from the isomorphism between the four-dimensional (4D) Lorentz group and those of the Mobius group for conformal transformations on the two-dimensional (2D) celestial sphere \cite{Strominger:2013jfa, Strominger:2014pwa, Kapec:2014opa,Kapec:2016jld, Strominger:2017zoo,Lam:2017ofc}. In the case of asymptotic plane wave states, the boundary conformal primary wavefunction generally follows from the Fourier transform of the bulk-to-boundary propagator defined on hyperbolic foliations of Minkowski spacetime. The massless limit is realized as a Mellin transform of plane waves, resulting in the energy dependence of bulk fields being traded for a scaling dimension dependence in the corresponding boundary operators. As a consequence, the resulting correlation functions of celestial conformal operators are manifestly $SL(2,\mathbb{C})$ invariant observables in a boost eigenbasis, which are proposed as duals of $S$-matrix elements in an energy-momentum eigenbasis \cite{Pasterski:2016qvg, Pasterski:2017kqt, Pasterski:2017ylz}. 

Celestial conformal field theories (CCFT)  have been investigated primarily from perturbative flat-spacetime amplitudes and are associated with infinite-dimensional asymptotic symmetry algebras~\cite{Guevara:2021abz, Strominger:2021lvk, Strominger:2021mtt, Atanasov:2021oyu}. Soft theorems for scattering amplitudes are realized through conformal soft theorems in CCFT ~\cite{Donnay:2018neh, Pate:2019mfs, Puhm:2019zbl}, that constrain the operator product expansions of celestial correlation functions~\cite{Pate:2019lpp}.  A remarkable property of CCFTs, on account of their involvement of boost scattering states, is that they invoke the entire energy spectrum of a theory and thus access their infrared (IR) and ultraviolet (UV) properties~\cite{Arkani-Hamed:2020gyp, McLoughlin:2022ljp}. CCFTs possess several properties similar to conformal field theories (CFT), including a conformal block expansion~\cite{Atanasov:2021oyu} and state-operator correspondence~\cite{Crawley:2021ivb}. However, due to their correspondence with scattering amplitudes on flat spacetime, they differ from CFTs in certain respects. This includes the presence of complex scaling dimensions for normalizable states and a delta function over the 2D cross ratio in CCFT correlation functions, with the latter due to the translation invariance of scattering amplitudes in momentum space~\cite{Arkani-Hamed:2020gyp, Atanasov:2021oyu, Pasterski:2021rjz, Raclariu:2021zjz}. More recent developments include investigations on CCFTs to leading loop orders~\cite{Albayrak:2020saa,Gonzalez:2020tpi,Donnay:2023kvm}, leading backreaction effects~\cite{Pasterski:2022lsl}
and their formulation on non-trivial asymptotically flat spacetimes~\cite{Pasterski:2020pdk, Gonzo:2022tjm}.  The celestial description of non-perturbative eikonal amplitudes was also initiated in~\cite{PipolodeGioia:2022exe}, which further demonstrates a correspondence with eikonal amplitudes in AdS/CFT~\cite{Cornalba:2007zb}. The relationship between CCFT and CFT$_2$ correlation functions have also been explored in~\cite{Casali:2022fro, Iacobacci:2022yjo, Sleight:2023ojm,deGioia:2023cbd}. We also note other potentially complementary approaches to holographic descriptions of scattering amplitudes on flat spacetime, which includes Carrolian holography defined along the codimension-one null boundary of flat spacetime~\cite{Donnay:2022aba, Donnay:2022wvx, Bagchi:2022emh, Bagchi:2023fbj, Saha:2023hsl, Saha:2023abr}, and derivations from the large AdS-radius limit in AdS/CFT~\cite{Hijano:2019qmi, Hijano:2020szl, Banerjee:2022oll, Duary:2022pyv, Duary:2022afn, Li:2023azu}

In this paper, we extend the analysis of~\cite{PipolodeGioia:2022exe} on eikonal amplitudes in flat spacetime to ones in the near-horizon region of a Schwarzschild black hole, with an impact parameter $x_{\perp}$ comparable to the Schwarzschild radius. This amplitude has been investigated in detail over recent years~\cite{tHooft:1996rdg, Gaddam:2020rxb, Gaddam:2020mwe, Betzios:2020xuj, Gaddam:2022pnb}. The motivation for such an amplitude can be traced back to eikonal amplitudes defined on flat spacetimes, which address trans-Planckian scattering processes with center of mass energies far greater than the Planck mass, i.e.,  $\sqrt{s} \gg M_{Pl}$ and correspondingly large impact parameters, i.e.,  $x_{\perp} \gg G_N \sqrt{s}$ with $G_N$ Newton's constant. On the other hand, flat spacetime eikonal amplitudes are expected to diverge in the regime of $x_{\perp} \sim G_N \sqrt{s}$ due to strong gravitational effects associated with the formation of a Schwarzschild black hole with a radius of $G_N \sqrt{s}$. This can be interpreted as an IR divergence associated with absorptive soft graviton exchanges, thereby reflecting the need for soft graviton bremsstrahlung \cite{Amati:1990xe} to yield IR-finite results in accordance with Weinberg's approach to IR divergences \cite{Weinberg:1965nx}. These properties motivate a possible eikonal description of the scattering process in the near-horizon region with a different kinematic regime, i.e., $x_{\perp} \sim G_N M$ with $M$ the mass of a background black hole, as considered in \cite{Gaddam:2020rxb, Gaddam:2020mwe, Betzios:2020xuj, Gaddam:2022pnb}. The resultant near-horizon eikonal amplitude then provides a description in the case of impact parameters comparable to a Schwarzschild radius, with the eikonal phase dominated by soft graviton exchanges.

This provides an interesting setting for a CCFT investigation on two grounds. First, since the near-horizon region of a Schwarzschild black hole around the bifurcation sphere can be well approximated by a flat spacetime frame in the small angle approximation, translation symmetry of massless particles can be restored in this frame. High energy massless states in an eikonal scattering process near the horizon can thus be investigated using known CCFT approaches on flat spacetime. The underlying global conformal symmetries of CCFTs originate from the isometries of flat spacetime. This then generalizes known CCFTs defined on flat spacetime to those defined only in the near-horizon region of a Schwarzschild black hole. Second, due to the gravitational effects of the background black hole that manifest in the phase of near-horizon eikonal amplitude, we expect the resulting CCFT to be quite different from those for scattering on flat spacetime. The detailed dynamics for the near-horizon eikonal scattering have been studied in the aforementioned works \cite{Gaddam:2020rxb, Gaddam:2020mwe, Betzios:2020xuj, Gaddam:2022pnb}. In a boost basis, the corresponding CCFT amplitudes can be expressed as the product of a universal conformal block for external-state conformal primaries with large conformal weights (inclusive of intermediate exchanges) and a conformally invariant function of cross-ratio $z=\frac{-t}{s}\ll 1$. The former provides a universal kinematic factor, while the latter captures the underlying dynamics of CCFTs or its parent theory in the momentum basis. 

As we will show,  the resultant eikonal phase obtained in \cite{Gaddam:2020rxb, Gaddam:2020mwe, Betzios:2020xuj,Gaddam:2022pnb} has a soft UV behavior, suggesting a possible UV completion in the near-horizon regime.
We find a closed-form result for near-horizon CCFT amplitudes, which to all loop orders, has a leading scaling behavior of
$z^{-1}$. This can be noted as being softer than those for celestial eikonal amplitudes on flat spacetime with massive mediating particles, which at tree-level scales like $\left(\sqrt{z}\right)^{-\beta}$ \cite{PipolodeGioia:2022exe}. Here, $\beta = \sum_{i=1}^4 \Delta_i - 4 \gg 1$, with $\Delta_i$ being the scaling dimensions of the external boosted eigenstates. 
More significantly, the near-horizon CCFT has poles at $\beta=-2n$ with $n \in \mathbf{N}$ labeling the loop order. Since the loop order corresponds to the number of exchanged soft gravitons in the ladder diagrams, this implies these poles are IR divergences due to the exchange of soft gravitons in the eikonal limit. Interestingly, the near-horizon CCFT is free from any poles for $\text{Re}\, \beta > 0$, as might be expected from a generic UV complete field theory, with an expansion for the amplitude $\sum_{n=0}^{\infty} a_n^{\text{UV}} \omega^{-2n}$. These results are further consistent with the observation of \cite{Arkani-Hamed:2020gyp, McLoughlin:2022ljp} that CCFT amplitudes for UV soft theories, such as those with a stringy Hagedorn spectrum, only have negative integer poles and correspond to the production of microscopic black holes \cite{Susskind:1993ws, Horowitz:1996nw, Lin:2007gi}. They may likewise be realized in a theory with only an IR soft expansion, with the amplitude going as $\sum_{n=0}^{\infty} a_n^{\rm IR} \omega^{2n}$. Thus, our results imply that the $\beta=-2n$ poles are more or less universal for strong gravity regimes, which can manifest either through black hole production or the existence of an event horizon.  

The rest of our paper is organized as follows. In the next section, we review the derivation of the 2D near-horizon eikonal amplitude from a perturbative analysis on the Schwarzschild background through a spherical harmonic decomposition of the fields. In section \ref{4dnhea}, we proceed to derive the 4D near-horizon celestial eikonal amplitude. We first uplift the 2D amplitude to a 4D partial sum amplitude in a near-horizon region about the bifurcation sphere. The spacetime considered is a nearly flat region that arises in a small angle and large black hole limit of the near-horizon background. We then carry out the partial resummation over small angles to derive a 4D momentum space eikonal amplitude. As this amplitude involves massless external states, following the prescription in~\cite{PipolodeGioia:2022exe}, we derive the near-horizon celestial eikonal amplitude from the Mellin transform. In section \ref{prop}, we study properties of the near-horizon celestial eikonal amplitude. This involves its exact evaluation to all loop orders. The CCFT result further provide IR poles and an overall $z$ dependence and we discuss their physical implications. We conclude with a discussion of our results and future directions in section \ref{end}.

\section{Review of 2D black hole Eikonal scattering amplitudes}

In this section, we provide a \emph{detailed review} of the near-horizon eikonal scattering amplitude considered in \cite{Gaddam:2020rxb, Gaddam:2020mwe, Betzios:2020xuj, Gaddam:2022pnb}. The conventional eikonal limit of trans-Planckian 2-2 scattering in flat spacetime, with a center-of-mass energy $\sqrt{s}$ far larger than the Planck mass $M_{PL}$ so that graviton exchanges dominate, requires a large impact parameter to suppress the transverse momentum transfer $q_{\perp}$. Additionally, to avoid divergent results caused by gravitational collapse near the scattering center, the impact parameter $x_{\perp} \sim {\cal O}(\hbar/q_{\perp})$ should also be far larger than the Schwarzschild radius associated with the center-of-mass energy, i.e., $x_{\perp}\gg G_N \sqrt{s}$ \cite{Veneziano:2004er, Amati:2007ak}.  The resultant eikonal amplitude (for massless particles) is \cite{tHooft:1987vrq, Amati:1987uf, Kabat:1992tb}
\be 
i{\cal M}=\frac{i \kappa^2 s^2}{q_{\perp}^2} \frac{\Gamma(1-i G_N s)}{\Gamma(1+i G_N s)} \left(\frac{4 \mu^2}{q_{\perp}^2}\right)^{-i G_N s}
\ee
with $\mu$ denoting an infrared cutoff. 
This amplitude has a semi-classical interpretation as a 1-1 scattering of an ultra-high energy massless particle against a null-like shockwave background, which incorporates the backreaction \cite{tHooft:1987vrq}. This is consistent with the expectation of an eikonal limit as a resummation over ladder graviton exchanges in a coherent background. Decomposing this eikonal amplitude in a partial wave basis yields a unitary S-matrix for each mode represented by an eikonal phase,
\be 
\label{flat-eikonal}
\delta_{\ell}(s) = \frac{s}{2} \log{\frac{\ell(\ell+1)}{s}}\;.
\ee
This phase encodes the peculiar dynamics from dominant soft graviton contributions in the ladder diagrams. Later, we will compare this phase to the one from eikonal scattering in the near-horizon region.

The metric of the near-horizon region of the Schwarzschild black hole is approximately a flat metric with an implicit horizon scale in relation to the Rindler metric. This raises the possibility of formulating eikonal scattering in the near-horizon region in a similar fashion to the approach on flat spacetime. Indeed, this idea had been proposed long ago \cite{Dray:1984ha, Dray:1985yt}, and has been recently refined with further details~\cite{Gaddam:2020rxb, Gaddam:2020mwe, Betzios:2020xuj, Gaddam:2022pnb, Feleppa:2023eoi}. Due to being restricted to the near-horizon region, the kinematic constraints for eikonal scattering are quite different from those on flat spacetime. This especially concerns the impact parameter, which is restricted to be $\ell_{PL}\ll x_{\perp} \approx R$, where $\ell_{PL}$ is the Planck length and $R=2 G_N M$ is the Schwarzschild radius. As shown in \cite{Gaddam:2020mwe, Gaddam:2020rxb}, eikonal scattering (small angle scattering) in this regime requires $\sqrt{s}\gg \gamma M_{PL}$ with $M_{PL}$ the Planck mass and $\gamma =\frac{M_{PL}}{M}$ an emerging dimensionless coupling between matter and gravitons of the effective theory that results from integrating out the transverse directions. Due to the smallness of $\gamma$ for a typical macroscopic black hole, the new constraint on $s$ implies that the eikonal scattering can be non-Planckian in the near-horizon region. Consequently, this enables us to circumvent the breakdown of conventional eikonal amplitudes when dealing with scattering at small impact parameters. 

\begin{figure}[h]
\begin{center}
\includegraphics[scale=0.5]{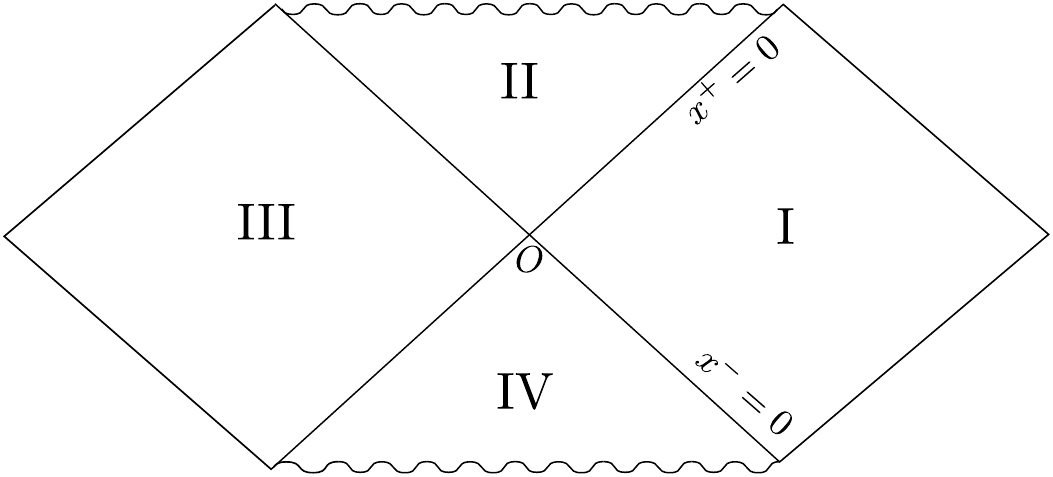}
\end{center}
\caption{Kruskal spacetime with regions I, ... IV; bifurcation sphere $O$ and $x^{\pm} = 0$ lines indicated.
}\label{fig1}
\end{figure}

The Schwarzschild spacetime has the following metric in static coordinates,
\begin{equation}
ds^2_{\text{Schwarzschild}} = - \left(1 - \frac{R}{r}\right) dt^2 + \left(1 - \frac{R}{r}\right)^{-1} dr^2 + r^2 d\Omega^2_2 
\label{sch.met}
\end{equation}
where $R = 2 G_N M$ is the Schwarzschild radius and $M$ is the black hole mass. To consider the near-horizon geometry, we perform a transformation to Kruskal coordinates, which is regular at the horizon and describes the maximally extended spacetime. This can be derived from the following definitions for $x^-$ and $x^+$, 
\begin{align}
x^- x^+ &= 2 R^2 \left(1 - \frac{r}{R}\right)  e^{\frac{r}{R}-1} \;; \notag\\
 \frac{x^-}{x^+} &= e^{\frac{t}{2R}} \qquad \text{Regions I and III}\,, \notag\\
 &= - e^{\frac{t}{2R}} \quad\; \text{Regions II and IV}\,,  \label{kruskgad}
\end{align}
with the event horizons located at $x^- x^+ = 0$ (Fig. \ref{fig1}).

This leads to \ref{sch.met} taking the form,
\begin{equation}
ds^2_{NH}=g_{\mu\nu}dx^{\mu}dx^{\nu}=-2A(x^-,x^+)dx^- dx^+ +r^2(x^-,x^+)d\Omega_2^2 \label{gad.met}\;,
\end{equation}
with 
\begin{equation}
    A(x^-,x^+)= \frac{R}{r(x^-,x^+)} e^{1-\frac{r(x^-,x^+)}{R}}\;.
\end{equation}
We will consider the theory of linearized Einstein gravity minimally coupled with a massless scalar $\psi$ in this background, 
\be
S[h_{\mu\nu},\psi]= \int d^4x\sqrt{-g} ~\Big[\frac{1}{4} R^{(1)}[h_{\mu\nu};g_{\mu\nu}] + \frac{1}{2}\psi~ \Box \psi +\frac{1}{2}\kappa h^{\mu\nu}T_{\mu\nu}\label{action} \Big]
\ee
where $\frac{1}{4}\sqrt{-g}~ R^{(1)}[h_{\mu\nu};g_{\mu\nu}]$ is the $h_{\mu\nu}$-quadratic part of $\frac{1}{2\kappa^2} \sqrt{-(g+\kappa h)} ~R[g_{\mu\nu}+\kappa h_{\mu\nu}]$, $\kappa=\sqrt{8\pi G_N}$, and $T_{\mu\nu}$ is the stress tensor of the scalar $\psi$
\begin{equation}    T_{\mu\nu}=\partial_{\mu}\psi \partial_{\nu}\psi-\frac{1}{2}g_{\mu\nu}g^{\rho\sigma}\partial_{\rho}\psi\partial_{\sigma}\psi \,.
\end{equation}

Exploiting the background spherical symmetry, one can decompose the metric and scalar fields in a spherical harmonic basis $Y^{m}_{\ell}$, i.e.
\begin{align}
h_{\mu\nu}=\sum_{\ell,m}h^{\text{odd}}_{\ell m,\mu\nu}Y^m_{\ell}+\sum_{\ell,m}h^{\text{even}}_{\ell m,\mu\nu}Y^m_{\ell}\;, \qquad \psi=\sum_{\ell,m} \psi_{\ell m}Y^m_{\ell}
\end{align}
with the additional choice of the usual Regge-Wheeler gauge{\footnote{We will suppress $\ell,m$ in scalar and graviton modes from this point onwards.}}, 
\begin{equation}    
h_{aA}^{\text{odd}}=-h_a \epsilon_A{}^B\partial_B Y^m_{\ell},\qquad h_{ab}^{\text{even}}=H_{ab}Y^m_{\ell},\qquad h_{AB}^{\text{even}}=K\gamma_{AB}Y^m_{\ell} 
\end{equation}
where $\gamma_{AB}$ and $\epsilon_{AB}$ are respectively the metric and Levi-Civita tensor on the 2-sphere, while $a,b$ are indices for the longitudinal null coordinates $x^{\pm}$. Spherical symmetry further ensures the decoupling of the even and odd modes. Moreover, the even parity nature of the scalar field yields no coupling to the odd modes from the interaction vertex. Therefore, only even parity graviton modes $K$ and $H_{ab}$ will be involved in the scattering of the massless scalar in the reduced theory. 

Since the longitudinal part of the near-horizon metric is conformal to a flat metric, we can perform the following Weyl transformation and field redefinitions to yield canonical kinetic terms in the reduced theory 
\be\label{Weyl}
g_{ab}\rightarrow A(x^-, x^+)\eta_{ab},\qquad H_{ab}\rightarrow \frac{1}{r}A(x^-,x^+){\mathbf{h}}_{ab},\qquad K\rightarrow \frac{1}{r}{\bf K},\qquad \psi \rightarrow \frac{1}{r}\phi\;.
\ee
This can be used to obtain a 2D effective theory by integrating out the transverse degrees of freedom. By introducing a single traceless tensor mode, $\tilde{\mathbf{h}}^{ab}= \mathbf{h}^{ab} - \eta^{ab}(\frac{1}{2} \mathbf{h} + {\bf K})$, for the 3-vertex coupling to two scalars, and carrying out a field redefinition $\tilde{{\bf K}}= {\bf h}+\frac{2R^2}{\ell(\ell+1)+2}\left(\eta^{ab} \partial_a \partial_b-\frac{1}{R^2}\ell(\ell+1)\right){\bf K}$, we can also remove the mixed contribution between ${\bf K}$ and $\mathbf{h}^{ab}$. The corresponding graviton propagators are complicated due to the potentials arising from the Weyl scaling and field redefinitions \ref{Weyl}. However, if we focus on the near-horizon region so that the metric becomes that of flat spacetime
\be 
A(x^-,x^+) \approx 1\;, \qquad \textrm{if} \quad r=R + {\cal O}(r-R)\;,
\label{snh}
\ee
the resulting 2D effective theory has considerably simpler properties.

Upon Fourier transforming all the fields and taking the $r \to R$ limit of \ref{snh}, the 2D effective action takes on the following simple form~\cite{Gaddam:2020rxb, Gaddam:2020mwe, Betzios:2020xuj, Gaddam:2022pnb}:
\bea
\label{effectiveaction}
S[\tilde{{\mathbf{h}}}^{ab}, \tilde{{\bf K}}, \phi] &=& \frac{1}{4} \int d^2k \; \Big( {\tilde{\mathbf{h}}}^{ab} {\bf P}^{-1}_{abcd}(k) \tilde{{\mathbf{h}}}^{cd}+ \tilde{{\bf K}}{\bf P}^{-1}_{\bf K} \tilde{\bf K} \Big) + \frac{1}{2} \int d^2p \; \phi {\bf P}^{-1}_{\phi} (p) \phi
\nn \\ 
&& + \; \gamma \int d^2\Pi\; 
\tilde{\mathbf{h}}^{ab}(k) p_{1a} p_{2b}  \phi_0(p_1) \phi(p_2)
\label{ftact}
\eea  
where $d^2\Pi$ is a shorthand for $d^2k d^2p_1 d^2 p_2 \; \delta^{(2)}(k+p_1+p_2)$, and the dimensionless coupling for the 3-vertex is given by 
\be
\gamma := \frac{\kappa}{R}= \frac{M_{PL}}{M}\;.
\ee
The propagators have the expressions
\begin{align}
{\bf P}_{\phi} (p) =\frac{1}{p^2+\mu^2-i\epsilon}\;, \qquad 
 {\bf P}_{\bf K} =\frac{4 R^2}{\ell (\ell +1)+2}\;,\qquad
{\bf P}^{abcd}(k) = {\bf P}^{abcd}_{\text{soft}}+ {\bf P}^{abcd}_{\text{hard}}(k)\;, \label{Pabcd}
\end{align}
where we decompose the tensor mode propagator into its soft ($k$-independent) and hard ($k$-dependent) parts as follows \cite{Gaddam:2022pnb}:
\begin{align}
&{\bf P}^{abcd}_{\text{soft}} = \frac{R^2}{\ell (\ell +1)+2} \left(\eta^{ab} \eta^{cd} - \eta^{ac}\eta^{bd} - \eta^{ad}\eta^{bc} \right)\;, \label{Pabcd_s}\\
& {\bf P}^{abcd}_{\text{hard}} (k) = -\frac{\ell (\ell +1)+2}{\ell (\ell +1)-2}\frac{1}{k^2+\mu^2}(\eta^{ab}+k^{ab})(\eta^{cd}+k^{cd}) \label{Pabcd_h}
\end{align}
with
\be
k^{ab}:=\frac{2 R^2}{\ell (\ell +1)+2}(k^a k^b-\frac{1}{2}k^2 \eta^{ab})\;.
\ee
The soft graviton exchange associated with ${\bf P}^{abcd}_{\text{soft}}$ will give a leading contribution to scattering amplitudes. While integrating out the transverse part by using the orthogonality relations between spherical harmonics, all the fields in the effective field theory acquire an effective mass 
\be
\mu^2 := \frac{\ell (\ell +1)+1}{R^2} \;,
\label{effm}
\ee
which can be thought of as an infrared regulator. 

We summarize an important assumption used in the derivation of the interaction vertex in \ref{ftact}, namely the absence of partial wave mixing. Apart from the interaction vertex, all other terms in the effective action up to quadratic order involve decoupled partial waves due to the spherical symmetry of the background. To preserve this property for the interaction vertex, it was argued in~\cite{Gaddam:2020rxb, Gaddam:2020mwe} that scattering processes that do not distribute angular momenta across the external legs through Clebsch-Gordan coefficients are those that preserve the background spherical symmetry. This can be satisfied by fixing one of the scalar particles in the interaction 3-vertex, $\phi_0$, to have no angular momentum (as in the last line of \ref{ftact}). 
We now elaborate more on this point. In general, the reduced action for the interaction vertex between the graviton and external scalars will, in general, involve exchanges of angular momenta so that it takes the following form
\begin{equation}
S_{\text{vertex}} = \frac{\gamma}{2} \sum_{\ell,m}\sum_{\ell_1,m_1}\sum_{\ell_2,m_2}\int d\Omega\, Y^m_{\ell}(\Omega) Y^{m_1}_{\ell_1}(\Omega) Y^{m_2}_{\ell_2}(\Omega) \int d^2x\, h^{ab}_{\ell m} \partial_a \phi_{\ell_1 m_1}  \partial_b \phi_{\ell_2 m_2} \,,
\label{s.vert}
\end{equation}
which evaluates to involve the sum over Clebsch-Gordan (CG) coefficients\footnote{The CG coupling is related to the CG coefficients $\langle \ell_1 , \ell_2 \,; m_1 \,, m_2 \vert \ell_1 , \ell_2 \,; \ell\,, m \rangle$ via \cite{Sakurai:1167961}
\begin{align}
&\int d\Omega\, Y^m_{\ell}(\Omega) Y^{m_1}_{\ell_1}(\Omega) Y^{m_2}_{\ell_2}(\Omega) \notag\\
&\qquad = \sqrt{\frac{(2 \ell_1 + 1)(2 \ell_1 + 1)}{4 \pi (2 \ell + 1)}} \langle \ell_1 , \ell_2 \,; 0 \,, 0 \vert \ell_1 , \ell_2 \,; \ell\,; 0 \rangle\langle \ell_1 , \ell_2 \,; m_1 \,, m_2 \vert \ell_1 , \ell_2 \,; \ell\,, m \rangle \;.
\label{3y}
\end{align}}. 
As a result, the general interaction vertex \ref{s.vert} involves partial wave mode mixings. In \cite{Gaddam:2020rxb, Gaddam:2020mwe}, it was argued that such mode mixings are associated with large transverse momenta exchanges that introduce non-spherical corrections of the background geometry. To suppress such non-spherical backreaction as the semi-classical approximation requires, we must lift the mixing of partial waves of different $\ell$ and $m$ in \ref{s.vert}.  
This can be implemented by fixing one of the external particles to be a s-wave ($\ell_2 = 0$ or $\ell_1 = 0$). This consequently simplifies the vertex action \ref{s.vert} to
\begin{equation}
S_{\text{vertex}} = \gamma \sum_{\ell,m} \int d^2x\, h^{ab}_{\ell m} \partial_a \phi_{\ell m}  \partial_b \phi_{0} \,,
\label{s.vert2}
\end{equation}
using $\int d\Omega\, Y^m_{\ell}(\Omega) Y^{m_1}_{\ell_1}(\Omega) = \delta_{\ell \ell_1} \delta_{m m_1}$, with the overall factor of $2$ accounting for either scalar particle being considered in the $\ell = 0$ state. The Fourier transform of \ref{s.vert2} is what appears as the interaction term in \ref{ftact}.

We also note that transverse exchanges are realized through the $\ell$ dependent effective mass term in \ref{effm}.  This will have a role in the description of external states and the eikonal approximation for the reduced theory in the following.
From the effective theory in \ref{effectiveaction}, we can obtain the Feynman rules to compute the amplitudes for the soft/hard graviton exchanges in the 2-2 scattering. In the vanishing effective mass limit, the external states are massless scalar particles described by longitudinal plane waves, with incoming momenta $p_1=(p_{1+},0)$ and $p_2=(0,p_{2-})$, which define the Mandelstam ${\bf s}$ (center-of-mass energy squared) in terms of 2D momenta as
\be 
{\bf s} =-(p_1 + p_2)^2= 2 p_{1+} p_{2-} \;.
\label{mand.s}
\ee

 Since only the tensor mode is coupled to the scalars via the interaction vertex, we just need to consider 2-2 scattering amplitude involving the exchange of tensor modes associated with graviton propagators ${\bf P}^{abcd}_{\text{hard}}$ and ${\bf P}^{abcd}_{\text{soft}}$. We denote the corresponding amplitudes as $M_{\rm hard}$ and $M_{\rm soft}$. Using the symmetry property: $k^{ab}=k^{ba}$ and $k^{ab}p_{1a} p_{2b} =0$, the hard graviton exchanges contribute to 
\be
M_{\rm hard}\propto ~\frac{\gamma^2 {\bf s}^2}{{\bf s}+\mu^2 }\;,
\ee
while from the soft graviton exchange one can obtain,
\be
M_{\rm soft}=(i\gamma p_{1a} p_{1b}) (2 {\bf P}^{abcd}_{\text{soft}}) (i\gamma p_{2c} p_{2d})=\frac{\gamma^2 R^2 {\bf s}^2}{\ell^2+\ell+2}\;.
\ee

Note that this 2D soft amplitude is suppressed for large $\ell$, contrary to the large $\ell$ dominance of 4D eikonal amplitude in flat space. Two important properties can be inferred from the above results. Due to the effective mass involving the Schwarzschild radius, we have a modified regime for eikonal scattering in the near-horizon region
\begin{equation}
{\bf s} \gg \mu^2 \quad \textrm{or} \quad \sqrt{{\bf s}} \gg \gamma M_{PL}\;.
\end{equation}

In addition, in the large ${\bf s}$ limit we always have 
\begin{equation}
\frac{M_{\rm hard}}{M_{\rm soft}} \sim  \mathcal{O}({\bf s}^{-1}) \,.
\end{equation}

Thus, $M_{\rm hard}$ is a subleading contribution to $M_{\rm soft}$ in the large ${\bf s}$ limit. As the soft graviton exchange dominates for all loop orders, one can re-sum the corresponding ladder diagrams to derive the leading contribution to the near-horizon eikonal amplitude \cite{Gaddam:2020mwe}
\be  
i A^{\ell}_{\rm NH \; eikonal}({\bf s})= 2 {\bf s} \left( e^{i \chi_{\ell}({\bf s})} - 1\right)\label{BHEA2}
\ee
where the associated eikonal phase is 
\be
\chi_{\ell}({\bf s}) = \frac{\kappa^2 {\bf s}}{\ell^2+\ell + 2}\;.
\label{nhep}
\ee
with $\kappa=\gamma R$. The ${\bf s}$ and $\ell$ dependences of this phase differ from the ones in eikonal amplitudes on flat spacetimes \ref{flat-eikonal}.  In the following section, we generalize the eikonal amplitude in \ref{BHEA2} 
 to a 4D amplitude \emph{within a small angle approximation near the bifurcation sphere} and subsequently provide its CCFT description in a boost eigenbasis.

\section{Construction of near-horizon celestial eikonal amplitude}\label{4dnhea}

We first provide detailed arguments for uplifting the 2D \emph{black hole eikonal amplitude} \ref{BHEA2} to a 4D partial wave amplitude in a near-horizon region considered for a large black hole in the small angle approximation. We then accordingly re-sum the 4D partial wave result using known techniques in the small angle approximation about flat spacetimes \cite{Levy:1969cr, Bautista:2021wfy} to derive an amplitude defined in terms of four-momenta and the impact parameter. The 4D eikonal amplitude which follows from the partial wave amplitude in the vanishing scattering angle limit, is hence only defined in a near-horizon Minkowski frame around the bifurcation sphere of the maximally extended Schwarzschild spacetime, which will be shown to be consistent with the approximations used in the derivation of the 2D near-horizon eikonal amplitude reviewed in the previous section. We lastly perform the Mellin transform on this momentum space amplitude to derive a celestial correlator on the horizon.

\subsection{Uplifting the partial wave eikonal amplitude to four dimensions}

To uplift \ref{BHEA2} to a 4D amplitude two key issues need to be addressed. The first concerns the kinematic constraint for the ``eikonal limit'' in the effective 2D theory, which will differ in a 4D spacetime. Thus, it is a priori unclear if one can promote the 2D black hole eikonal amplitude to a 4D one, which can also allow for a CCFT description. For general amplitudes, such kinematic lifting could be difficult to realize. Thus, we need to consider a particular set of amplitudes for our purpose. 
On a related note, we would also need to address the status of momentum conservation for scattering involving 4D momenta since translation is, in general, broken on black hole spacetimes. The second issue concerns the possible mixing between partial wave modes due to introducing transverse exchanges. In the presence of partial wave mode mixing, the resummation of eikonal amplitudes will need to generalize the contribution from the interaction vertex to the eikonal phase in \ref{BHEA2}. Since both issues are closely associated with the possible difference in kinematic symmetries between Minkowski and Schwarzschild spacetimes, the best way to resolve them is to discuss the recovery of Minkowski isometries in the near horizon region. In the following, we argue that this can be achieved in a small angle and large black hole radius approximation, and this is of particular relevance for near-horizon amplitudes satisfying the eikonal approximation and spherical symmetry in the Minkowski frame, as we now explain.

We will be interested in the leading contribution of the metric \ref{gad.met} in the $r \to R$ limit  \footnote{One may also be interested in the near-horizon metric up to ${\cal O}(r-R)$,
\begin{equation*}
ds^2 = - 2 dx^- dx^+ + R^2  d\Omega^2_2 + \left[4 \left(\frac{r}{R}-1\right)  dx^- dx^+ +2R^2\left(\frac{r}{R}-1\right)d\Omega^2_2\right]+\cdots\;. \label{nh.met1}
\end{equation*}}
\begin{equation}
ds^2_{\text{NH}} = - 2 dx^- dx^+ + R^2  d\Omega^2_2 + \mathcal{O}(R^{-1}) \,.\label{nh.met}
\end{equation}

In further considering a small angle approximation, i.e. considering the leading planar approximation to the angular coordinates in a region far smaller than $R$, the spacetime can be transformed to a flat spacetime metric, noted as a `Minkowski coordinate frame' in~\cite{tHooft:1996rdg}, around the bifurcation sphere in the maximally extended Schwarzschild spacetime. More specifically if we consider $d\Omega^2_2 = d \theta^2 + \sin^2 \theta d \phi^2$, and assume that the transverse directions $X$ and $Y$ are related to the angles $\theta$ and $\phi$ via \cite{tHooft:1996rdg} \footnote{If instead we considered $d\Omega^2_2 = \frac{4}{(1 + z \bar{z})^2} dz d\bar{z}$, then the transformations $z = \frac{X}{2 R} + i \frac{Y}{2 R}$ and $\bar{z} = \frac{X}{2 R} - i \frac{Y}{2 R}$ would also recover the flat spacetime metric}
\begin{align}
X = R \left(\theta - \frac{\pi}{2}\right)\;, \quad & \quad Y = R \phi
\label{nh.fmt2}
\end{align}
\ref{nh.met} gives the Minkowski coordinate frame metric 
\begin{equation}
ds^2_{\text{Minkowski frame}} = -dx^{+}dx^{-} + dx_{\perp}^2\,.
\label{met.mf}
\end{equation}
where we have replaced $R^2 d \Omega_2^2 = dX^2 + dY^2 := dx_{\perp}^2$ \footnote{We have rescaled $x^{\pm}\rightarrow x^{\pm}/\sqrt{2}$ for convenience.}. The small angle approximation is implemented above by considering the leading order terms in a Taylor series expansion of the transverse metric with $\theta$ small \footnote{More specifically, we have $\sin \theta = \sin \left(\frac{X}{R} + \frac{\pi}{2}\right) = 1 + \mathcal{O}\left(\frac{X}{R}\right)$}. Therefore, Mikowski isometries are formally recovered in this approximation.  While translation invariance is generally broken on a black hole spacetime, it follows from the isometries of \ref{met.mf} that translation invariance and, consequently, momentum conservation are satisfied by scattering processes within the Minkowski coordinate frame. We stress that the Mikowski coordinate frame metric \ref{met.mf} for the near horizon geometry holds exactly only in the large $R$ limit, so that the subleading ${\cal O}(R^{-1})$ terms are negligible. An important consequence of the large $R$ limit is that we have a geometry with large transverse directions. As a consequence, we may consider forward scattering processes with small transverse momenta exchanges in the Minkowski coordinate frame, which we consider in the following. 

Based on the Minkowski frame metric in the large $R$ limit of the near-horizon geometry, which is nothing but the Minkowski space, we may directly apply the formalism of CCFT construction for flat spacetime to our near-horizon eikonal amplitude by performing a Mellin transformation. However, we note that 2D kinematic variables, such as  Mandelstam variables, differ from the 4D ones by lack of transverse directions. For general scattering states, following \ref{met.mf} the 4D Mandelstam variable $s$ in the Minkowski frame will be given by, 
\be
s = -(p_1 + p_2)^2 \approx 2 p_1^+ p_2^- - p_{\perp}^2\;.
\ee
This is incompatible with the 2D Mandelstam ${\bf s}$ variable given in \ref{mand.s} for generic $p_{\perp}^2 \ne 0$. To bypass this difficulty, we will restrict our consideration to \emph{forward scattering amplitudes with small transverse momenta exchange}. Therefore for the 2-2 scattering process we define,
\begin{align}
p_i^{+} &= p_i^0+p_i^3 \gg p_{i, \perp} \,, \qquad p_i^- \simeq 0  \quad \text{for}\,\quad  i= 1,3 \;; \notag\\
p_i^- &= p_i^0 - p_i^3 \gg p_{i, \perp} \,,\qquad p_i^+ \simeq 0  \quad \text{for}\,\quad  i= 2,4\;; \notag\\
t& = - (p_1 + p_3)^2\,, \qquad s = -(p_1 + p_2)^2 \approx 2 p_1^+ p_2^- \approx {\bf s} \,,
\label{ps4d}
\end{align}
where $i=1,2$ labels the incoming particles, and $i=3,4$ the outgoing particles. Our consideration of $p_{i,\perp}$ small is consistent with $p_A\simeq 0$ as adopted in \cite{Gaddam:2020rxb, Gaddam:2020mwe, Betzios:2020xuj, Gaddam:2022pnb}, and such states can also be realized naturally in the context of trans-Planckian scattering \cite{Dray:1984ha, Verlinde:1991iu, Verlinde:1993mi}.

In short, to promote the 2D kinematic relations to the 4D ones in the near-horizon Minkowski coordinate frame, and with the purpose of subsequently deriving a CCFT description, we will only consider forward scattering amplitudes for the 2-2 process with external states satisfying \ref{ps4d}. This subset of amplitudes can be lifted from 2D to 4D while respecting eikonal kinematics. For more general scattering amplitudes, the kinematic lifting will be more nontrivial.

 We will now provide the upliftment of the near-horizon eikonal amplitude. A general 4D N-particle scattering amplitude can be formally obtained from the partial wave analysis as follows 
\be
A^N_{4D}={\cal N} \sum_{\{\ell_i,m_i\}} \prod_i Y_{\ell_i,m_i} (\hat{p}_i) A^{\{\ell_i,m_i\}}_{\rm p.w.}
\label{mode-sums}
\ee
where $i=1,\cdots, N$ label the external particles with momenta $p_i$ (and $\hat{p}_i$ denote their orientation), ${\cal N}$ is a normalization constant, $Y_{\ell_i,m_i}(\hat{p_i})$'s are the spherical harmonics, and $A^{\{\ell_i,m_i\}}_{\rm p.w.}$ is the partial wave amplitude equipped with the constraints of (angular)-momentum conservation. The above formal sum generally leads to a complicated kernel for transforming the partial wave amplitudes to the corresponding 4D amplitude. However, as we have discussed, for our purpose of constructing the CCFT dual of uplifted eikonal amplitudes in the near-horizon regime, we will only consider forward scattering semi-classical amplitudes. Due to small transverse momenta exchanges in these amplitudes, additional partial wave mode mixings are not introduced. This is consistent with the absence of partial wave mode mixing to prevent non-spherical backreaction in the semi-classical analysis of \cite{Gaddam:2020rxb, Gaddam:2020mwe, Betzios:2020xuj, Gaddam:2022pnb}. For the 2-2 scattering, this reduces the multi-sums over $(\ell_i,m_i)$ into a single sum of the overall $(\ell,m)$, i.e., it is reflected in the fact that the partial wave amplitude \ref{BHEA2} depends only on the overall $(\ell,m)$. We further recall that the label $\ell$ in \ref{BHEA2} refers to the partial wave of one of the external states in the 2-2 process, with the other particle fixed to be a $\ell = 0$ state. Thus, by the aforementioned assumption\footnote{The sum over $(\ell,m)$ can be further simplified by the formula
\be 
\sum_{m = -\ell}^{\ell} Y_{\ell m}(\hat{p}) Y^*_{\ell m}(\hat{p}') = \frac{2 \ell + 1}{4 \pi} P_{\ell} (\cos \theta)
\ee
because the 2-2 partial wave amplitude $A^{\ell}_{\rm NH \; eikonal}$ is $m$-independent.
}, the above formal partial wave summation for the 2-2 eikonal scattering in the near-horizon Minkowski frame simplifies to \cite{DiVecchia:2023frv, Muzinich:1987in}
\begin{align}
A_{\text{NH eikonal}} = \frac{\mathcal{N}}{4 \pi} \sum_{\ell = 0}^{\infty} (2 \ell + 1) P_{\ell} (\cos \theta) \; A^{\ell}_{\rm NH \; eikonal} (s)
\label{TA}
\end{align}
where $A^{\ell}_{\rm NH \; eikonal} (s) $ is given by \ref{BHEA2} with the argument in terms of the 2D Mandelstam ${\bf s}$ now replaced with the 4D Mandelstam $s$ (following \ref{ps4d}), and $\cos \theta = \hat{p}\cdot \hat{p}'$.  Note that our assumption of small transverse exchanges implies that \ref{TA} holds for small-angle scattering. Setting $\mathcal{N} = 4 \pi$ in \ref{TA}, we arrive at our final expression for the Minkowski frame partial wave amplitude
\begin{equation}
A_{\text{NH eikonal}} = 2 s \sum_{\ell} (2\ell + 1) \left[ \exp \left( \frac{i \kappa^2 s}{\ell^2+\ell+2}\right) - 1 \right] P_{\ell} (\cos \theta)\,.
\label{amp.pw}
\end{equation}
The normalization has been chosen to provide an overall scaling consistent with graviton mediated eikonal amplitudes and does not affect our analysis to follow.

In summary, the 4D amplitude $A_{\text{NH eikonal}}$ given in \ref{amp.pw} provides a consistent uplifting of the 2D partial wave eikonal amplitude given in \ref{BHEA2}, defined in the Minkowski frame about the bifurcation sphere. We caution the reader that this does not apply for general scattering amplitudes in the near-horizon Minkowski frame. It works only for semi-classical forward eikonal amplitudes, which is what we will consider for the dual CCFT description. \ref{amp.pw} bears the usual form for the eikonal scattering with $\chi_l = \frac{\kappa^2 s}{\ell^2+\ell + 2}$  playing the role of 4D eikonal phase. However, it differs from its counterpart \ref{flat-eikonal} for eikonal scattering on flat spacetime due to different underlying dynamics. While we kinematically go over to a flat spacetime Minkowski frame in the small angle and large Schwarzschild radius limit, the phase (resulting from non-vanishing curvature contributions in this limit) provides a different eikonal resummation than \ref{flat-eikonal} and captures near-horizon effects of the Schwarzschild spacetime on the scattering. In the following subsection, we proceed to evaluate \ref{amp.pw} using known techniques in flat spacetime for small angle scattering.

\begin{figure}[h]
\begin{center}
\includegraphics[scale=0.3]{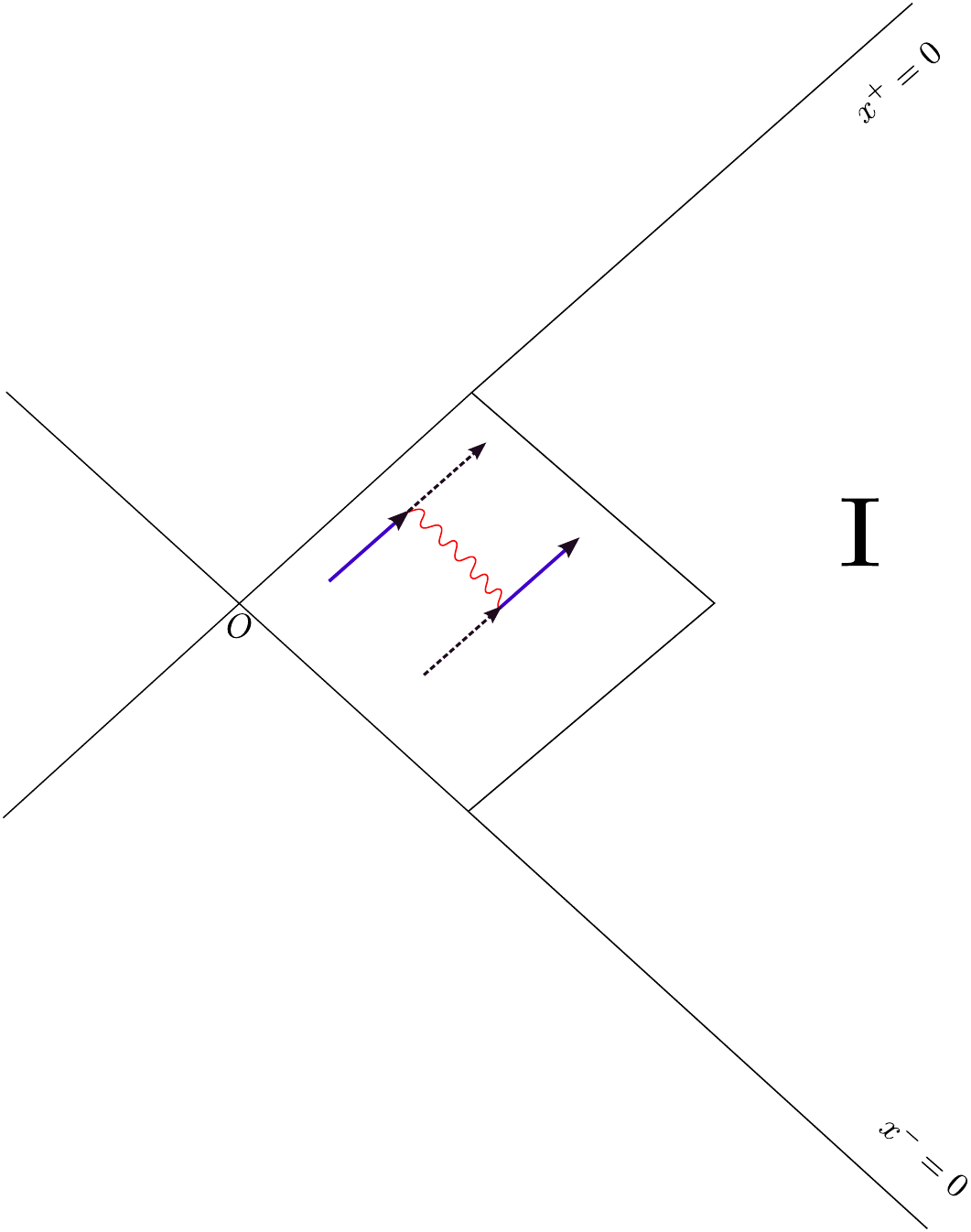}
\end{center}
\caption{Minkowski frame defined in an $\mathcal{O}(R)$ region about the bifurcation sphere $O$ in the exterior region I of the global Kruskal spacetime. Also indicated is a representative 2-2 scattering process between particle with label $\ell$ (bold) and $\ell = 0$ (dashed) mediated by a soft graviton (wavy).
}\label{fig2}
\end{figure}

As the near-horizon eikonal amplitude is a high energy forward scattering process involving massless plane waves as external states, a dual celestial description can be derived using the Mellin transform on the external states. We will return to a more detailed discussion of these properties in Sec.~\ref{NHCA:sec}. For the moment, we point out three key differences with flat-spacetime CCFT constructions:
\begin{itemize}
\item Time is rescaled by a factor of $\frac{1}{2 R}$ (and a constant) relative to the `global time coordinate' $t$ in \ref{sch.met}. As such, while we will still denote the frequency as $\omega$ in the Mellin transform, it is related to the frequency at null infinity by a factor of $2 R$.

\item As evident from \ref{nh.met}, the asymptotic conformal boundary of the spacetime is entirely a portion of the past and future event horizons about the bifurcation sphere and not null infinity. 
\item Dual celestial correlators will be constructed only for eikonal scattering processes (with small transverse momentum exchange) respecting background spherical symmetry in the \emph{Minkowski frame} (see Fig.\ref{fig2}) near the bifurcation sphere.
\end{itemize}

We also note there exist earlier holographic proposals based on AdS$_3$ spacelike foliations of the near-horizon geometry in the limit of approaching the non-degenerate horizon \cite{Sachs:2001qb, Gibbons:2008hb} which realize the above features. In particular, the Schwarzschild spacetime is conformal to an optical metric with time rescaled by $\frac{1}{2 R}$ and with the conformal boundary located at the event horizon. However, this formalism provides no particular advantage over known approaches in flat spacetime for investigating scattering processes in the near-horizon region involving plane wave states.

\subsection{4D near-horizon eikonal amplitude from partial sum}

In this subsection we will carry out the explicit sum of \ref{amp.pw}. We recall that in \ref{amp.pw} $\ell$ labels the total angular momentum for one of the partial waves in the scattering amplitude, $s$ is the square of the center of mass energy in 4D momentum space, and $\theta$ is the small angle between the incoming and scattered particles. The sum over $\ell$ in \ref{amp.pw} can be traded for a 2D integral over the transverse directions. This follows from the relation between the transverse direction and the angular momentum mode $\ell$ arising from the definition of angular momentum (squared) \cite{Verlinde:1991iu, Verlinde:1993mi}
\begin{equation}
\ell(\ell+1) = s \vert x_{\perp}\vert^2\;.
\label{lx.rel}
\end{equation}
This turns the eikonal phase $\chi_{\ell}$ given in \ref{amp.pw} to 
\begin{align}
\chi_{\ell} =  \frac{i \kappa^2 s}{s \vert x_{\perp}\vert^2 + 2} =  \frac{ \kappa^2 }{ \vert x_{\perp}\vert^2} + \mathcal{O}(s^{-1})\;,
\label{eik.se}
\end{align}
which receives its dominant contribution for large $\ell$ (or equivalently large $s$) and small $x_{\perp}$, and is different from the corresponding graviton-mediated eikonal amplitude on flat spacetime that grows with $s$~\cite{Kabat:1992tb, Verlinde:1991iu}
In the following, we retain the complete expression for the eikonal phase in \ref{eik.se}, which captures all $\mathcal{O}(s^{-1})$ properties. However, as in all eikonal amplitudes, all corrections in the more subleading $\mathcal{O}\left(\sqrt{-\frac{t}{s}}\right)$ are ignored.

Another ingredient is an integral representation of the Legendre polynomials for small-angle scattering, for which we have
\begin{align}
  P_{\ell}(\cos{\theta})&=  \frac{1}{2\pi}\int_0^{2\pi}d\phi~\exp\left(i \,2 \ell \sin\left(\frac{\theta}{2}\right)\cos{\phi}\right)\;,  \label{leg.rel}\\
  \sin\left(\frac{\theta}{2}\right) & = \frac{\sqrt{-t}}{2 \sqrt{s}} = \frac{\vert p_{\perp}\vert}{2 \sqrt{s}} \,,\label{mand.sa}
\end{align}
where we used the known relation between Legendre polynomials and Bessel functions in the small angle approximation in \ref{leg.rel}, while \ref{mand.sa} is the relation between the exchanged momentum and center of mass energy for small angle scattering.

Substituting \ref{lx.rel}, \ref{leg.rel} and \ref{mand.sa} in \ref{amp.pw}, we find the following result for the 4D near-horizon eikonal amplitude (by ignoring all subleading $\mathcal{O}\left(\sqrt{-\frac{t}{s}}\right)$ corrections)
\begin{align}
A_{\text{NH eikonal}} = 2 s \int d^2 x_{\perp} \left[\exp\left( \frac{i \kappa^2 s}{s \vert x_{\perp}\vert^2 + 2}\right) - 1 \right] e^{i \vec{p}_{\perp}.\vec{x}_{\perp}} \;.
\label{BHEA}
\end{align}
This differs from the expression for the eikonal amplitude on asymptotically flat spacetimes through the eikonal phase. The two main differences lie in the dependence of the eikonal phase on $s$ and the impact parameter $x_{\perp}$. The eikonal phase for graviton-mediated scattering on asymptotically flat spacetimes grows with $s$ and holds for large impact parameters $x_{\perp}$, which is evident from substituting \ref{flat-eikonal} in \ref{lx.rel}~\cite{Verlinde:1991iu}. In contrast, the subleading terms in the eikonal phase of the near-horizon scattering process decay with large $s$ (as indicated in \ref{eik.se}) with the dominant contribution from $x_{\perp} \ll 1$ in units of the Schwarzschild radius. In other words, one can see from \ref{eik.se} that while we consider a high energy scattering process with $\frac{-t}{s} \ll 1$ near the horizon, the eikonal phase grows more dominant as we reduce the impact parameter.  
In the following subsection, we determine how this manifests in a celestial description and compare our result with the celestial eikonal amplitude on asymptotically flat spacetimes. 

\subsection{Near-horizon celestial eikonal amplitude}\label{NHCA:sec}

We have noted that the near-horizon geometry in the limit of approaching the horizon can be described by a flat spacetime metric. The near-horizon eikonal amplitude is a scattering process restricted to this region involving external massless plane wave states. Hence a near-horizon celestial description of this amplitude can result from a Mellin transform of the near-horizon eikonal amplitude through its action on the external states, following the same arguments as recently provided for flat spacetime eikonal amplitudes in \cite{PipolodeGioia:2022exe}. 
 
We accordingly define the 4D 2-2 near-horizon celestial eikonal amplitude as the Mellin transform of the near-horizon eikonal amplitude given in \ref{BHEA} including the momentum conserving delta function
\begin{align}
    \tilde{A}_{\text{NH eikonal}} =(2\pi)^4 \left( \prod_{i=1}^4\int_0^{\infty} d\omega_i \omega_i^{\Delta_i-1}\right) A_{\text{NH eikonal}}~\delta^{(4)}\left(\sum_{i=1}^4 p_i\right) \,.
    \label{mel.nhe}
\end{align}

We consider the momenta of the external states as in~\cite{PipolodeGioia:2022exe}. This involves an all-outgoing convention
\begin{equation}
p_i = \eta_i \omega_i \hat{q}_i \;;\qquad i= 1 \,, \cdots 4
\label{mom.param}
\end{equation}
with $\eta_i = + 1$ for the outgoing states $i=3,4$, $\eta_i = - 1$ for the incoming states $i=1,2$, and $\hat{q}_i$ a null vector parameterized in terms of longitudinal and transverse components $(q_i^0\,, q_{i,\perp}\,, q_i^3)$ as 
\begin{align}
\hat{q}_i &= \left(1 + q_i\,, q_{i,\perp}\,, 1-  q_i\right) = (1+z_i \bar{z}_i, z_i+\bar{z}_i, -i(z_i-\bar{z}_i), 1-z_i\bar{z}_i) \,, \qquad i=1,3 \label{q.out1}\\
\hat{q}_i &= \left(1 + q_i\,, q_{i,\perp}\,, -1+  q_i\right) = \frac{1}{z_i\bar{z}_i}  (1+z_i \bar{z}_i, z_i+\bar{z}_i, -i(z_i-\bar{z}_i), 1-z_i\bar{z}_i)\,, \qquad i=2,4 \,. \label{q.out2}
\end{align}
where ($z, \bar{z}$) is a point on the celestial sphere at the horizon. The massless condition $\hat{q}^2_i = 0$ imposes $4 q_i = \vert q_{i,\perp} \vert^2$, which relates the longitudinal $q_i$ with the transverse two component vector $q_{i,\perp}$. The constraints then turn into the expressions of $\hat{q}_i$'s in terms of $z_i$'s.  Hence for an eikonal scattering process with $s \gg - t$, we have $q_i \sim \vert q_{i,\perp} \vert^2 \ll 1$, with $\omega_1 \simeq \omega_3$ and $\omega_2 \simeq \omega_4$. We use $\simeq$ to indicate an equivalence up to corrections subleading in $\mathcal{O}\left(q_{\perp}^2\right)$. 

The above considerations for the external states further imply that for $i=1,3$ we have $p_i^+ = p_i^0+p_i^3=2 \eta_i \omega_i \gg p_{i, \perp}$ and $p_i^- =p_i^0-p_i^3\simeq 0$, while for  $i= 2,4$ we have $p_i^- = 2 \eta_i \omega_i \gg p_{i, \perp}$ and $p_i^+ \simeq 0$, as in \ref{ps4d}. With the above definitions, the delta function in \ref{mel.nhe} takes the form
\begin{align}
\delta^{(4)}\left(\sum_{i=1}^4 p_i\right) & = 2 \delta\left(p_1^+ + p_3^+\right) \delta\left(p_2^- + p_4^-\right) \delta^{(2)}\left(\sum_{i=1}^4 p_{i,\perp}\right)\;, \notag\\
& = \frac{1}{2} \delta\left(\omega_1 - \omega_3\right) \delta\left(\omega_2 - \omega_4 \right) \delta^{(2)}\left(\sum_{i=1}^4 \eta_i \omega_i q_{i,\perp}\right) \;.
\label{delta.fac}
\end{align}

Likewise, for the Mandelstam variables $s$ and $t$ defined in \ref{ps4d} we have
\begin{align}
s & \simeq - 2 p_1^{+} p_2^{-} = 4 \omega_1 \omega_2 \;,  \notag\\
-t & \simeq \left( p_{1,\perp} + p_{3,\perp}\right)^2 = \left( \omega_3 q_{3,\perp} - \omega_1 q_{1,\perp} \right)^2 \;.
\label{mand.con}
\end{align}

Substituting \ref{delta.fac} and \ref{mand.con} in \ref{mel.nhe}, we get the following expression
\begin{align}
    \tilde{A}_{\text{NH Eikonal}} &=4(2\pi)^4  \int_0^{\infty} d\omega_1 \int_0^{\infty} d\omega_2\, \omega_1^{\Delta_1+\Delta_3 -1 }\omega_2^{\Delta_2+\Delta_4 -1 } \notag\\
    & \qquad  \int d^2x_{\perp} \sum_{n=1}^{\infty}\frac{1}{n!}\left( \frac{i \kappa^2 }{ \vert x_{\perp}\vert^2 + \frac{1}{2\omega_1\omega_2}}\right)^n e^{-i\omega_1 q_{13,\perp}.x_{\perp}} \delta^{(2)}(\omega_1 q_{13,\perp}+\omega_2 q_{24,\perp})\,,
    \label{mel.nhe2}
\end{align}
where we have written the eikonal phase as a series expansion and used the notation $q_{i,\perp} - q_{j,\perp} = q_{ij,\perp}$.

With the exponential representation for the 2D delta function
\begin{equation}
\int d^2 \bar{x}_{\perp} \exp\left(- i k\cdot \bar{x}_{\perp}\right) = (2\pi)^2 \delta^{(2)}(k)\,,
\label{delta.def}
\end{equation}
we may express \ref{mel.nhe2} as
\begin{align}
    \tilde{A}_{\text{NH Eikonal}} &=4(2\pi)^2  \int_0^{\infty} d\omega_1 \int_0^{\infty} d\omega_2\, \omega_1^{\Delta_1+\Delta_3 -1 }\omega_2^{\Delta_2+\Delta_4 -1 } \notag\\
    & \qquad  \int d^2x_{\perp} \int d^2\bar{x}_{\perp} \sum_{n=1}^{\infty}\frac{1}{n!}\left( \frac{i \kappa^2 }{ \vert x_{\perp}\vert^2 + \frac{1}{2\omega_1\omega_2}}\right)^n e^{-i\omega_1 q_{13,\perp}.\left(x_{\perp} + \bar{x}_{\perp}\right)} e^{-i\omega_2 q_{24,\perp}. \bar{x}_{\perp}} \,.
    \label{mel.nhe3}
\end{align}
The second line in \ref{mel.nhe3} can be further simplified. We first express all frequencies appearing in the eikonal phase as resulting from the action of celestial momentum operators on $\omega_i^{\Delta_i - 1}$ for the incoming particles, $P_i^{\mu} = - \hat{q}_i^{\mu} e^{\partial_{\Delta_i}}$ for $i=1,2$ \cite{Stieberger:2018onx}. We can expand the eikonal phase as an infinite series
\be 
 \left(\vert x_{\perp}\vert^2 + \frac{1}{2\omega_1\omega_2}\right)^{-n} = \sum_{k=0}^{\infty} \; C_{n-1}^{k+n-1}  x_{\perp}^{-2(n+k)} \left(-\frac{1}{2\omega_1\omega_2}\right)^{k}
\ee
with $C_{n-1}^{k+n-1} = \frac{(k+n-1)!}{k! (n-1)!}$.
From the standard identity of the celestial momentum operator acting on the integrand of the Mellin transform \cite{PipolodeGioia:2022exe,Stieberger:2018onx}
\begin{align}\label{der}
\omega_1^{\Delta_1+\Delta_3-1 -k} & = e^{ - k\partial_{\Delta_1}} \omega_1^{\Delta_1+\Delta_3-1}\;, \notag\\
\omega_2^{\Delta_2+\Delta_4-1-k } & = e^{ - k\partial_{\Delta_2}} \omega_2^{\Delta_2+\Delta_4-1}\;,
\end{align}
we have the relation
\begin{align}
\left( \frac{i \kappa^2 }{ \vert x_{\perp}\vert^2 + \frac{1}{2\omega_1\omega_2}}\right)^n & \omega_1^{\Delta_1+\Delta_3 -1 }\omega_2^{\Delta_2+\Delta_4 -1 }\notag\\
&= \left( \frac{i \kappa^2 }{ \vert x_{\perp}\vert^2 + \frac{1}{2} \exp(-\partial_{\Delta_1})\exp(-\partial_{\Delta_2}) }\right)^n \omega_1^{\Delta_1+\Delta_3 -1 }\omega_2^{\Delta_2+\Delta_4 -1} \;.
\label{del.der}
\end{align}

We hence see that a factor of $s = 4 \omega_1 \omega_2$ can be interpreted as arising from the action of a shifting operator $4 e^{\partial_{\Delta_1}} e^{\partial_{\Delta_2}}$ in the celestial basis following \ref{der}.
We in addition consider the transformation $x_{\perp} + \bar{x}_{\perp} \to x_{\perp}$ resulting in an eikonal phase that depends on the transverse distance $x_{\perp} - \bar{x}_{\perp}$. By performing this transformation and using \ref{del.der} in \ref{mel.nhe3}, we get the result
\begin{align}
    &\tilde{A}_{\text{NH Eikonal}} =4(2\pi)^2  \int d^2x_{\perp} \int d^2\bar{x}_{\perp} \sum_{n=1}^{\infty}\frac{1}{n!}\left( \frac{i \kappa^2 }{ \vert x_{\perp} - \bar{x}_{\perp}\vert^2 + \frac{1}{2} \exp(-\partial_{\Delta_1})\exp(-\partial_{\Delta_2}) }\right)^n \notag\\ 
    & \qquad\qquad \qquad  \int_0^{\infty} d\omega_1  \omega_1^{\Delta_1+\Delta_3 -1 } e^{-i\omega_1 q_{13,\perp}. x_{\perp}} \int_0^{\infty} d\omega_2  \omega_2^{\Delta_2+\Delta_4 -1} e^{-i\omega_2 q_{24,\perp}. \bar{x}_{\perp}}    \notag\\
    & \qquad := 4(2\pi)^2  \int d^2x_{\perp} \int d^2\bar{x}_{\perp} \left(e^{i \hat{\chi}_{\rm NH}} - 1\right) \frac{i^{\Delta_1 + \Delta_3} \Gamma\left(\Delta_1 + \Delta_3\right)}{\left(-q_{13,\perp}\cdot x_{\perp} + i \epsilon\right)^{\Delta_1 + \Delta_3}}\frac{i^{\Delta_2 + \Delta_4} \Gamma\left(\Delta_2 + \Delta_4\right)}{\left(-q_{24,\perp}\cdot \bar{x}_{\perp} + i \epsilon\right)^{\Delta_2 + \Delta_4}} \,,
    \label{mel.nhe4}
\end{align}
where in the last equation we defined the eikonal phase operator 
\begin{equation}
\hat{\chi}_{\rm NH} := \frac{\kappa^2}{ \vert x_{\perp} - \bar{x}_{\perp}\vert^2 + \frac{1}{2} \exp(-\partial_{\Delta_1})\exp(-\partial_{\Delta_2})}\;,
\label{gf.gadd}
\end{equation}
and made use of the identity
\begin{equation}
\int_0^{\infty} d\omega  \omega^{\Delta -1 } e^{-i \eta\omega q. x_{\perp}} = \frac{i^{\Delta} \Gamma\left(\Delta\right)}{\left(- q \cdot x_{\perp} + i \eta \epsilon\right)^{\Delta}}\;.
\label{cpm.def}
\end{equation}

\ref{cpm.def} provides a definition of massless conformal primary wavefunctions for scattering on asymptotically flat spacetimes, while the form of \ref{mel.nhe4} is very similar to the result for flat spacetime celestial eikonal amplitudes given in ~\cite{PipolodeGioia:2022exe}. The main difference with the flat spacetime result stems from the form of the eikonal phase operator~\ref{gf.gadd}. In \cite{PipolodeGioia:2022exe} the eikonal phase operator for graviton mediated scattering takes the following form \footnote{The form of \ref{chi_flat} is consistent with \ref{flat-eikonal} after replacing the action of shift operators by $s=4\omega_1\omega_2$, along with the relation \ref{lx.rel} and the explicit form of $G_{\perp}(x_{\perp},\bar{x}_{\perp}) \sim \ln|x_{\perp}-\bar{x}_{\perp}|$.}
\be 
\hat{\chi}_{\rm flat}\propto  \kappa^2 \exp(\partial_{\Delta_1})\exp(\partial_{\Delta_2}) G_{\perp}(x_{\perp},\bar{x}_{\perp})
\label{chi_flat}
\ee
where $G_{\perp}(x_{\perp},\bar{x}_{\perp})$ is the transverse part of the propagator of exchange graviton. Comparing \ref{gf.gadd} and \ref{chi_flat}, we see that one difference concerns the action of $\exp(\partial_{\Delta_1})\exp(\partial_{\Delta_2})$, with the inverse dependence appearing in near horizon eikonal amplitudes.  In the following section, we will investigate \ref{mel.nhe4} in more detail and compare our results with those of celestial eikonal amplitudes on asymptotically flat spacetimes.

\section{Properties of the near-horizon celestial eikonal amplitude}\label{prop}

Celestial amplitudes, on account of their involvement of boost eigenstates, 
are known to be sensitive to both UV and IR properties of scattering processes \cite{Arkani-Hamed:2020gyp}. Lorentz and translation invariance on the celestial sphere further manifest in certain universal properties that celestial amplitudes must possess. 

The near-horizon celestial eikonal amplitude \ref{mel.nhe4} can be expanded as the sum of the Feynmann ladder diagrams, which are classified by the number of exchanged soft gravitons, i.e., denoted by $n$, or $(n-1)$ loops, 
\begin{align}
\tilde{A}_{\text{NH eikonal}} = \sum_{n=1}^{\infty} \tilde{A}_{\text{NH eikonal}}^{(n)} \label{An.NHt}
\end{align}
with 
\begin{align}
& \tilde{A}_{\text{NH eikonal}}^{(n)} =  \frac{4(2\pi)^2}{n!} \int d^2x_{\perp} \int d^2\bar{x}_{\perp} \left(i\hat{\chi}_{\rm NH} \right)^n
  \frac{i^{\Delta_1 + \Delta_3} \Gamma\left(\Delta_1 + \Delta_3\right)}{\left(-q_{13,\perp}\cdot x_{\perp} + i \epsilon\right)^{\Delta_1 + \Delta_3}}\frac{i^{\Delta_2 + \Delta_4} \Gamma\left(\Delta_2 + \Delta_4\right)}{\left(-q_{24,\perp}\cdot \bar{x}_{\perp} + i \epsilon\right)^{\Delta_2 + \Delta_4}}\,.
 \label{mel.tchan0}
 \end{align}  
The form of \ref{mel.tchan0} is exact for all $n$. Below, we will argue that our result satisfies the expected universal properties of celestial amplitudes and is consistent with the defining properties of near-horizon eikonal amplitudes -- that is it is mediated by soft graviton exchanges in the large $s$ limit.

We start with the integral representation of celestial eikonal amplitude from \ref{mel.nhe2}, incorporating the definition \ref{An.NHt}
\begin{align}
   \tilde{A}_{\text{NH eikonal}}^{(n)} = &4(2\pi)^4\int_0^{\infty} d\omega_1 \omega_1^{\Delta_1+\Delta_3-1}\int_0^{\infty}d\omega_2 \omega_2^{\Delta_2+\Delta_4-1}\notag\\& \qquad \int d^2x_{\perp} \frac{1}{n!}\left(\frac{i\kappa^2}{\vert x_{\perp}\vert^2 + \frac{1}{2\omega_1\omega_2}}\right)^n e^{-i\omega_1 q_{13,\perp} \cdot x_{\perp}}\delta^{(2)}(\omega_1 q_{13,\perp}+\omega_2 q_{24,\perp})  \;. \label{mel.tchan}
\end{align}

One of the properties of this celestial eikonal amplitude is that the integration over $x_{\perp}$ can be evaluated exactly to give the modified Bessel function of the second kind of integer order $n-1$, 
\begin{align}
\int d^2 x_{\perp}\left(\frac{i \kappa^2}{\vert x_{\perp}\vert^2 + \frac{1}{2\omega_1\omega_2}} \right)^n e^{-i\omega_1 q_{13,\perp}\cdot x_{\perp}}  &=  2\pi \frac{i^n \kappa^{2n}}{(n-1)!}\left(\omega_1 \vert q_{13,\perp} \vert \sqrt{\frac{\omega_1 \omega_2}{2}}\right)^{n-1} \notag\\&\qquad K_{n-1}\left(\sqrt{\frac{\omega_1}{2\omega_2}} \vert q_{13,\perp}\vert \right)  
\label{Kn}
\end{align}

where we have used the known integral (cf 6.565.4 of \cite{Gradshteyn:1943cpj})

\begin{align}
    \int_0^{\infty}\frac{J_{\nu}(bx)x^{\nu+1}}{(x^2+a^2)^{\mu+1}}dx=\frac{a^{\nu-\mu}b^{\mu}}{2^{\mu}\Gamma(\mu+1)}K_{\nu-\mu}(ab),\quad -1<\text{Re}\, \nu<\text{Re}\,(2\mu+3/2); a,b>0\;.
\end{align}

Our analysis will be further considered in a center-of-mass frame with
\be
\vert q_{\perp}\vert:= \vert q_{13,\perp} \vert = \vert q_{24,\perp} \vert\;,
\ee
Hence, on substituting \ref{Kn} in \ref{mel.tchan} and performing the rescalings $\omega_1 \vert q_{\perp}\vert \to \omega_1$ and $\omega_2 \vert q_{\perp}\vert \to \omega_2$, we find
 \begin{align}
   \tilde{A}_{\text{NH eikonal}}^{(n)} = &4(2\pi)^5 \frac{i^n \kappa^{2n}}{n! (n-1)!} \left(\vert q_{\perp}\vert\right)^{-\beta-n-3}\int_0^{\infty} d\omega_1 \omega_1^{\Delta_1+\Delta_3-1}\int_0^{\infty}d\omega_2 \omega_2^{\Delta_2+\Delta_4-1}\notag\\& \left(\omega_1 \sqrt{\frac{\omega_1 \omega_2}{2}}\right)^{n-1} K_{n-1}\left(\sqrt{\frac{\omega_1}{2 \omega_2}} \vert q_{\perp}\vert \right)\delta^{(2)}(\omega_1 n_{13,\perp}+\omega_2 n_{24,\perp})\;,\label{mel.tchan2}
\end{align}
where we defined the two dimensional vectors
\begin{align}
 n_{13,\perp}=(n_{13,\perp}^1,  n_{13,\perp}^2):= \frac{q_{13,\perp}}{\vert q_{\perp}\vert }\;, \qquad  n_{24,\perp}=(n_{24,\perp}^1,  n_{24,\perp}^2):= \frac{q_{24,\perp}}{\vert q_{\perp}\vert }\;,
\label{q1234.norm}
\end{align}
and
\be
\beta := \Delta_1+\Delta_2+\Delta_3+\Delta_4 - 4.
\ee

To further simplify \ref{mel.tchan2}, we consider the 2-2 scattering in the center of mass frame with the following parametrization for the transverse momenta in terms of the cross-ratio $z= \frac{-t}{s}$~\cite{PipolodeGioia:2022exe},
\begin{align}
    q_{13,\perp}=(-(\sqrt{z}+\sqrt{\bar{z}}),i(\sqrt{z}-\sqrt{\bar{z}})) \;, & \quad q_{24,\perp}=(\sqrt{z}+\sqrt{\bar{z}}, i(\sqrt{z}-\sqrt{\bar{z}})) \label{COM_frame} \;,
\end{align}
so that
\be
\vert q_{\perp}\vert= 2\sqrt{\vert z\vert}\;, \qquad  n_{24,\perp}^1 n_{13,\perp}^2  = i \frac{z - \bar{z}}{\vert q_{\perp}\vert^2} = - n_{24,\perp}^2 n_{13,\perp}^1 \;. \label{q1234.prod}
\ee
In the parametrization of $\hat{q}_i$ of \ref{q.out1} and \ref{q.out2}, this corresponds to the choice
\be  
z_1=0, \quad z_2=\infty, \quad z_3=\sqrt{z}, \quad z_4=-\frac{1}{\sqrt{z}}\;,
\ee
such that in the eikonal limit,
\be
z= \frac{-t}{s} \approx \frac{z_{13}z_{24}}{z_{12}z_{34}} \ll 1\;.
\ee 

The 2D delta function $\delta^{(2)}(\omega_1 n_{13,\perp}+\omega_2 n_{24,\perp})$ can be factored into a product of delta functions \cite{PipolodeGioia:2022exe}
\begin{align}
   \delta^{(2)}(\omega_1 n_{13,\perp}+\omega_2 n_{24,\perp})&=\frac{1}{\omega_1}\delta\left(\omega_2+\omega_1\frac{n_{24,\perp}^1}{n_{13,\perp}^1}\right)\delta(n_{24,\perp}^1 n_{13,\perp}^2 -n_{24,\perp}^2 n_{13,\perp}^1)\;, \notag\\
   &= \frac{\vert q_{\perp} \vert^2}{2 \omega_1}\delta\left(\omega_2 - \omega_1\right) \delta(z - \bar{z}) \;,
    \label{2d.id}
\end{align}
where in the second line of \ref{2d.id} we made use of \ref{COM_frame} and \ref{q1234.prod}, along with the standard scaling property of the Dirac delta function. On carrying out the integral over $\omega_2$ in \ref{mel.tchan2} and using $\delta(z - \bar{z})$ to write $\vert q_{\perp}\vert = 2 \sqrt{z}$, we find
\begin{align}
 \tilde{A}_{\text{NH eikonal}}^{(n)} = & 2 (2\pi)^5 \frac{i^n \kappa^{2n} 2^{-\beta - 2}}{n! (n-1)!} 2^{\frac{3}{2}(1-n)} K_{n-1}(\sqrt{2 z}) \left(\sqrt{z}\right)^{-\beta-n-1} \delta(z - \bar{z}) \int_0^{\infty} d\omega_1 \omega_1^{\beta + 2n - 1}\;.\label{mel.tchan3} 
\end{align}
Lastly, we can perform the $\omega_1$ integral by analytically continuing to the regime of large scaling dimension $\Delta_i \gg 1$ with the following result \cite{Donnay:2020guq}
\begin{align}
\int_0^{\infty} d\omega_1 \omega_1^{\beta + 2n - 1}:= \bm{\delta}(i(\beta + 2n)) \;.
\label{del.temp}
\end{align}
Thus the integral in \ref{del.temp} is well defined when we set $\beta=-2n+ib$. 
Hence, on substituting \ref{del.temp} in \ref{mel.tchan3} we arrive at the final result

\begin{align}
 \tilde{A}_{\text{NH eikonal}}^{(n)} = &2 (2\pi)^6 \frac{i^n \kappa^{2n} 2^{-\beta - 2}}{n! (n-1)!} 2^{\frac{3}{2}(1-n)} K_{n-1}(\sqrt{2 z}) \left(\sqrt{z}\right)^{-\beta-n-1} \delta(z - \bar{z})\bm{\delta}(i(\beta + 2n)) \;.\label{mel.tchanf} 
\end{align}

The contribution $2^{-\beta - 2} \delta(z - \bar{z})$ in \ref{mel.tchanf} is part of the universal kinematic contribution in all celestial amplitudes \cite{Atanasov:2021cje, PipolodeGioia:2022exe}. The delta function over $z$ in particular manifests translation invariance on the celestial sphere. The scaling behavior of the amplitude will follow from $\left(\sqrt{z}\right)^{-\beta-n-1} K_{n-1}(\sqrt{2 z})$, which as we now show, has a universal $n$-independent leading scaling behavior for all  $\tilde{A}_{\text{NH eikonal}}^{(n\ge 2)}$.  The leading contribution of $K_{n-1}(\sqrt{2 z})$ at each order $n$ is given by
\begin{align}
K_{n-1}(\sqrt{2 z}) &\approx -\frac{1}{2} (-\sqrt{2 z})^{n-1}\partial_{z}^{n-1} \ln \left(\frac{z}{2}\right)\;.
\label{bk.exp}
\end{align} 

We consider \ref{bk.exp} in \ref{mel.tchanf} at tree level ($n=1$) and higher loops more generally ($n \ge 2$). For the tree-level exchange, we find
\begin{align}
 \tilde{A}_{\text{NH eikonal}}^{(1)} &=  - (2\pi)^6 i \kappa^{2} 2^{-\beta - 2} \ln \left(\frac{z}{2}\right) \left(\sqrt{z}\right)^{-\beta-2} \delta(z - \bar{z})\bm{\delta}(i(\beta + 2)) + \cdots\,, \notag\\
&= - (2\pi)^6 i \kappa^{2} \ln\left(\frac{z}{2}\right) \delta(z - \bar{z})\bm{\delta}(i(\beta + 2)) + \mathcal{O}\left(\sqrt{z}\right)\,,
 \label{mel.te} 
\end{align} 
where we utilized $\bm{\delta}(i(\beta + 2))$ in the second line of \ref{mel.te} and $\cdots$ in the first line represent contributions subleading in small $z$. The tree-level exchange is hence dominated by $z^0 \ln z$. Similarly, for $n\ge 2$ by using \ref{bk.exp} and $\bm{\delta}(i(\beta + 2n))$ in \ref{mel.tchanf}, we find the result to be 
\begin{align}
 \tilde{A}_{\text{NH eikonal}}^{(n \ge 2)} &=  (2\pi)^6 \frac{i^n \kappa^{2n}}{(n-1) n!} 2^{-\beta - (n+3)} \left(\sqrt{z}\right)^{-\beta-2(n+1)} \delta(z - \bar{z})\bm{\delta}(i(\beta + 2n)) + \cdots\,, \notag\\
&= (2\pi)^6 \frac{i^n \kappa^{2n}}{(n-1) n!} 2^{n - 3} z^{-1} \delta(z - \bar{z})\bm{\delta}(i(\beta + 2n)) + \mathcal{O}\left(\left(\sqrt{z}\right)^{-1}\right)
 \label{mel.le} \;.
\end{align} 
We hence find the universal leading behavior of $z^{-1}$ for all $n\ge 2$ loop orders in the near-horizon celestial eikonal amplitude. It is quite interesting to consider the dynamical interpretation of the factor $\bm{\delta}(i(\beta + 2n))$ in \ref{mel.tchanf}. Recall that the same pole structures are also proposed and discussed for the celestial amplitudes of IR or UV soft theories~\cite{Arkani-Hamed:2020gyp}. For the IR soft ones, these poles appear after Mellin transforming the IR expansion of the amplitude in the momentum basis, i.e., $\sum_{n=0}^{\infty} a_n^{\rm IR} \omega^{2n}$. In a UV soft theory, such as string theory, the UV behavior is softened by the productions of Hagedorn stringy states, which can be understood as the microscopic black holes by the string/black hole correspondence \cite{Susskind:1993ws, Horowitz:1996nw, Lin:2007gi}. This implies that the celestial amplitudes can capture both non-perturbative physics in both UV and IR sides through the characteristics of $\beta=-2n$ poles. In our case, we are considering the celestial eikonal amplitudes for which the $\beta=-2n$ appears with $n$ labeling the order of the loop/ladder diagrams. Thus, a natural interpretation of these poles is the dominance of the soft graviton exchanges, with $n$ corresponding to the number of soft gravitons appearing in the ladder diagrams. An additional consequence is the absence of poles for $\text{Re}\, \beta > 0$ in a manner analogous to UV soft theories. In this sense, these poles are the manifestation of the IR divergence due to soft graviton exchanges in the near-horizon region. This implies that celestial amplitudes can capture non-perturbative effects of strong gravity due to either black hole production or the existence of an event horizon.

\section{Summary and Conclusion}\label{end}
Eikonal amplitudes provide an important class of non-perturbative scattering processes that have been recently investigated in the celestial basis. In this paper, we considered the celestial description of eikonal amplitudes past the critical length scale for the impact parameter, which is taken to be large in the usual eikonal approximation. In this regime, the leading approximation to the eikonal amplitude is governed by a resummation over soft graviton exchanges in the near-horizon geometry of a Schwarzschild black hole. Hence, near-horizon celestial eikonal amplitudes provide an interesting case of non-perturbative scattering processes in the boost eigenbasis beyond those on flat spacetime.

The near-horizon eikonal amplitude~\cite{Gaddam:2020rxb, Gaddam:2020mwe, Betzios:2020xuj, Gaddam:2022pnb} accounts for the leading backreaction about a Schwarzschild background and is a 2D result following the integration over spherical harmonics. The resulting eikonal phase \ref{nhep} is dominated by small $\ell$ modes and transverse directions $x_{\perp}$. In section~\ref{4dnhea} we first uplifted the 2D near-horizon amplitude to a 4D amplitude under specific assumptions. We assumed a small angle and large black hole approximation to recover a planar flat spacetime region about the bifurcation sphere, that is invariant under Minkowski isometries and has a part of the future and past horizons as asymptotic boundaries. The 4D planar region further has large transverse directions, which allowed us to consider forward scattering processes that well approximate the kinematics and respect the interactions involved in the 2D near-horizon eikonal amplitude. As a result, the 4D near-horizon eikonal amplitude follows from a partial sum over the 2D amplitude, with the Mandelstam variables now appropriately in terms of 4D momenta.  We subsequently performed a Mellin transform on the massless external states to derive the near-horizon celestial eikonal amplitude.  The conformal primary wavefunctions describing the external states are the same as those for the celestial eikonal amplitude on flat spacetimes, which follows from the isometries of the near-horizon region in the small angle approximation being identical to those on flat spacetimes. However, the eikonal phase, which captures the interactions of the external states with the exchanged gravitons near the horizon, crucially differs from the flat spacetime eikonal phase. More specifically, the 4D near-horizon eikonal amplitude is defined from impact parameters comparable to the Schwarzschild radius and a perturbative series in $s^{-1}$ around $s$ being infinite. This manifests the property that the near-horizon eikonal amplitude is mediated by soft graviton modes.

In section~\ref{prop} we investigated the near-horizon celestial eikonal amplitude to derive the main results of our paper, namely an exact all-loop order result for the celestial amplitude. The result involves universal kinematic factors of celestial amplitudes on flat spacetime. This is expected from our consideration of the near-horizon region in the small angle approximation, which simply provides a flat spacetime for the scattering process. However, the dynamical content of the celestial amplitude differs considerably from celestial amplitudes on flat spacetime. One of these differences follows from the $\bm{\delta}(i(\beta + 2n))$ contribution. While this behavior is consistent with the expectation of soft UV behavior in CCFT, the near horizon celestial eikonal amplitude provides the specific representation of exchanged soft gravitons with loop order $n$. We, in addition, have a $\sqrt{z}^{-\beta - n - 1} K_{n-1}(\sqrt{2 z})$ term as an all-loop order contribution from the near-horizon eikonal phase. This term on expanding about $z \ll 1$, and in conjunction with $\bm{\delta}(i(\beta + 2n))$, provides a universal leading $n$-independent behaviour for $\sqrt{z}$. Thus, the leading scaling behavior of the cross-ratio $z$ in the celestial amplitude result is independent of loop order.

There are several further avenues to explore in the context of near-horizon amplitudes. It will be interesting to go beyond the small angle approximation used for the near-horizon background. In particular, we expect a correction to the $2$-sphere part of the metric in \ref{sch.met}, which are also known to influence near-horizon symmetries \cite{Donnay:2015abr, Donnay:2016ejv, Grumiller:2019fmp}. We expect the consideration of near-horizon eikonal amplitudes past the leading approximation at large $R$ to be useful in understanding the manifestation of near-horizon symmetries in celestial amplitudes. In addition, it will also be important to better understand the celestial correspondence between the 1-1 scattering of a massless scalar field on a shockwave background and the 2-2 graviton-mediated eikonal amplitude for massless external scalar fields. We expect this correspondence to hold for the near horizon eikonal amplitude upon considering a shockwave in the near horizon region.  \vspace{1em}

\paragraph{Acknowledgement :} 
The work of KF is supported by Taiwan's NSTC with grant numbers 111-2811-M-003-005 and 112-2811-M-003 -003-MY3. The work of FLL is supported by Taiwan's NSTC with grant numbers 109-2112-M-003-007-MY3 and 112-2112-M-003-006-MY3. The work of AM is supported by the Ministry of Education, Science, and Technology (NRF- 2021R1A2C1006453) of the National Research Foundation of Korea (NRF). \vspace{1em}

\bibliographystyle{jhep}

\bibliography{refs.bib}	
\end{document}